\documentclass[twocolumn,pra,showpacs,superscriptaddress]{revtex4-1}
\usepackage{latexsym,amssymb,amsmath,graphicx,bm,epstopdf}
\usepackage[dvipsnames]{xcolor}

\renewcommand\vec[1]{\mathbf{#1}}
\newcommand\cbdot{\bm{\cdot}}
\newcommand\mat[1]{\mathbf{#1}}
\newcommand\ket[1]{\left|#1\right>}

\begin{document}
\title{Bayesian recursive data pattern tomography}

\author{Alexander Mikhalychev}
\affiliation{Institute of Physics, Belarus National Academy of Sciences, F.Skarina Ave. 68, Minsk 220072 Belarus}
\author{Dmitri Mogilevtsev}
\affiliation{Institute of Physics, Belarus National Academy of Sciences, F.Skarina Ave. 68, Minsk 220072 Belarus}
\affiliation{Centro de Ci{\^e}ncias Naturais e Humanas, Universidade Federal do ABC, Santo Andr{\'e},  SP, 09210-170 Brazil}
\author{Yong Siah Teo}
\affiliation{Department of Optics, Palack{\'y} University, 17. listopadu 12, 77146 Olomouc, Czech Republic}
\author{Jaroslav {\v R}eh{\'a}{\v c}ek}
\affiliation{Department of Optics, Palack{\'y} University, 17. listopadu 12, 77146 Olomouc, Czech Republic}
\author{Zden{\v e}k Hradil}
\affiliation{Department of Optics, Palack{\'y} University, 17. listopadu 12, 77146 Olomouc, Czech Republic}

\begin{abstract}

We present a simple and efficient Bayesian recursive algorithm
for the data-pattern scheme for quantum state reconstruction, which is applicable to situations where measurement settings can be controllably varied efficiently. The
algorithm predicts the best measurements required to accurately reconstruct the
unknown signal state in terms of a fixed set of probe states.
In each iterative step, this algorithm seeks the measurement setting that minimizes
the variance of the data-pattern estimator, which essentially measures the reconstruction accuracy, with the help of a data-pattern bank that was acquired prior to the signal reconstruction. We show that with this algorithm, it is
possible to minimize the number of measurement settings required to obtain a reasonably accurate state estimator by using just the optimal settings and, at the same time, increasing the numerical efficiency of the data-pattern reconstruction.

\end{abstract}

\date{\today}

\maketitle

\section{Introduction}

In very general terms, quantum tomography involves a comprehensive
toolset that allows an observer to carry out verification and
diagnostics on quantum degrees of freedom. With it, the observer
is able to acquire maximal information about a quantum source by
characterizing its quantum state (quantum state tomography \mbox{\cite{QST1,QST2,QST3}}). To
perform this task, one usually needs a rather precisely calibrated
measurement set-up. Calibrating a detection device for signals on
an intensity level of a few quanta can be a challenge. In general,
measurement schemes are lossy and noisy. For example, the
so-called single-photon detectors are only able to distinguish
between the presence and absence of a signal pulse (hence the
alternative names ``bucket'' or ``on/off'' detectors). In
addition, these detectors have wavelength-dependent efficiencies
that are less-than-unity, and give dark photon counts in the
presence of noise that may originate from either thermal
fluctuation, afterpulsing for heterostructural detectors and other
sources. The dead times incurred from minimizing such spurious
counts further complicate photodetection. Precise calibration of
measurement apparatuses (quantum process or detector tomography)
is yet another well-established area in quantum tomography. More
sophisticated device-calibration methods would generally entail
the utilization of intrinsic quantum resources such as
entanglement.  Examples of such approaches include the ``absolute
calibration'' method {\cite{klyshko80,sergienko81}}, and
``self-testing'' or ``blind tomography''
{\cite{scarani,Navascues}}. Quite often, device calibrations are
carried out up to some experimental error uncertainties that are
difficult to control. There are situations where detector
calibration is impractical. For example, in wavefront sensing
experiments \cite{wavefront}, wavefront detectors that contain
microlenses are practically impossible to calibrate reliably as
they are, for close proximities between pairs of microlenses would
result in calibration interference.

In this article, we discuss a particular tomography procedure that
permits one to circumvent the need for measurement-device
calibration when reconstructing the quantum state of any arbitrary
signal --- the data pattern tomography procedure proposed in 2010
\cite{data patterns 2010}. The idea is somewhat similar to that of
optical image analysis with known optical response function
\cite{classic}. An observer measures responses (the data patterns,
so to speak) for a set of known quantum probe states. Then, the
observer matches them with the response obtained from the unknown
signal of interest. This method is naturally insensitive to
imperfections of the measurement set-up, since all device
imperfections are automatically incorporated into and accounted
for by the measured data patterns. This procedure can also be
understood, in terms of reconstruction-subspace optimization, as a
search for the optimal state estimator over the subspace that is
spanned by the probe states.

The data-pattern scheme with coherent states as probe states was
recently realized by the experimental groups in Oxford
\cite{oxford} and Paderborn \cite{paderborn} for the
reconstruction of few-photon states (thermal, heralded
single-photon and two-photon states). It was shown that the
data-pattern scheme is indeed feasible and is able to perform
robust and accurate state reconstruction. At present, however, the
procedure of data pattern inference is still far from optimal. In
experiments \cite{oxford,paderborn} the reconstruction was done by
considering only a set of probe states that was deemed
sufficiently large (48 in Ref.~\cite{oxford}, 50-150 in
Ref.~\cite{paderborn}). One should understand that the choices of
these ``sufficiently large'' numbers are not entirely
\textit{ad hoc}, as these choices are not based solely on the
average number of photons in the signal, which can be
inaccurate.

It is possible to estimate this ``sufficiently large number'' fairly accurately by making an educated guess of
a class of plausible signal states that would closely describe the
quantum source and choose a probe-state basis that best fits this
class of states. In the language of statistics, this is equivalent
to developing a useful \emph{prior} for the data-pattern
experiment. If the observer, after some rough preliminary
calibration of the source, has reasons to believe that the unknown
signal state is very likely residing in some operator subspace, he
can make use of this insight to define the set of probe states
that spans this subspace \cite{data patterns 2013,oxford,spiral}
 for data-pattern reconstruction \cite{data patterns 2010,data patterns 2013}.

From this pre-selected set of probe states and some fixed choice
of measurement settings (or measurements in short), it is
possible to optimize the inference procedure by incorporating
more probe states from this search space to enhance the
data-pattern fit. In this sense, a new and improved data-pattern
fit is obtained by acquiring some more data patterns and
incorporating them to the previous fit. However, if the
optimization of the data-pattern fit is restricted to a fixed set
of measurements, it may turn out that many of the pre-chosen probe
states play little role in enhancing the fit. One such example
would be the case of single-photon state inference in
Ref.~\cite{data patterns 2013}. In this report, approximately
a third of the measured probe states practically give no
constructive contributions to the final data-pattern fit in terms of reconstruction accuracy.

In the light of this finding, we propose an alternative
methodology to optimize the data-pattern scheme. For
situations in which adjusting measurement settings can be done
quite precisely and efficiently in a well-controlled manner, we
consider ways to optimally choose measurements for the signal
state to provide the best fit for the given set of probe states.
Since the probe states that collectively give a faithful
description of the signal state can always be reasonably
restricted to a finite number by some physical considerations of
the source that allows the observer to truncate its dimension,
one can optimize the subsequent measurements for a pre-chosen
set of probe states with the aim of optimally utilizing
these data patterns in order to increase the reconstruction
accuracy.

Prior to the signal reconstruction, data patterns for the probe
states are measured for various measurement settings and the
subset of optimal settings for the signal reconstruction are
directly determined from these patterns with a recursive numerical
algorithm. In this way, the number of patterns to be used for
signal reconstruction is minimized and numerical reconstruction
can hence be carried out efficiently. With this same bank of data
patterns, the observer can perform the same measurement
optimization for other signal states for which these pre-chosen
probe states are appropriate for their reconstruction.

For this purpose, we shall develop an efficient recursive
Bayesian iterative procedure for the measurement optimization.
The observer, with access to the data-pattern bank, can make use
of this procedure to generate a sequence of optimal measurement
settings that minimizes the reconstruction accuracy, and finally
terminate this sequence after a natural stopping criterion is met.
As a rule, when the reconstruction accuracy of the state
estimator becomes comparable with the improvement of the
data-pattern fit, it is then no longer necessary to perform any
more measurement. Incorporating more data patterns from any other
measurement will not lead to any appreciable increase in accuracy
\cite{rosett,enk,moroder,mogilevtsev2013}.

The outline of the article is as follows. In
Sec.~\ref{basics}, the basic elements of
the recursive procedure is introduced. Next, we describe
approximations that are valid for the data-pattern scheme for
coping with the complexities of calculating Bayesian integrals and
derive the expressions for the Bayesian update that is employed
in the procedure in Sec.~\ref{practicalities}. Following
which, in Sec.~\ref{choice}, we shall formally discuss the measurement optimization procedure based on the
Bayesian-update equations that were established in
Sec.~\ref{practicalities}, and its stopping criterion. Last,
but not the least, in Sec.~\ref{examples}, the mechanisms and
results of this Bayesian numerical procedure are illustrated with
some examples of low-intensity signal states with the help of
probe coherent states.

\section{Basics}
\label{basics}

To lay the groundwork for subsequent discussions, let us outline the general scheme of things. We assume that
there is a finite set of available probe states
described by the density operators $\rho_m$, where $m=1 \ldots M$. We would like to reconstruct the true signal state described by the density
operator $\rho$, which is known, to some level of confidence after a preliminary calibration of the source, to be representable as a linear combination
of these probe states,
\begin{equation}
\label{eqn:density_matrix}
\rho = \sum_{m=1}^M c^{\text{(true)}}_m \rho_m,
\end{equation}
where $c^{\text{(true)}}_m$ are real coefficients. The aim of the data-pattern inference
procedure is to optimize these coefficients. To this end, we
suppose that the experimental set-up allows us to vary measurement settings in a controlled and efficient manner, thereby allowing us to sequentially carry out different measurements on the source to obtain data patterns. Out of a total number of $K$ measurement settings employed, the $k$th setting yields one of two outcomes described by the positive operators
$\Pi_k$ and $1-\Pi_k$ that compose a positive operator-valued
measure (POVM) for this measurement. The probability of the outcome $\Pi_k$ for the
$k$th measurement performed on the signal state is given by Born's rule as
\begin{equation}
\label{eqn:elementary_prob}
P_k= \mathrm{Tr}\{\Pi_k\rho\}=\sum_{m=1}^M c^{\text{(true)}}_m p_{km}=\sum_{m=1}^M c^{\text{(true)}}_m\mathrm{Tr}\{\Pi_k\rho_m\}
\end{equation}
where $p_{km}$ is the probability for the measurement outcome
$\Pi_k$ and the $m$th probe state. In practice, we have
access to a finite number of signal and probe copies, say, $N$
each for every measurement. So, instead of probabilities $P_k$
and $p_{km}$, one measures frequencies $F_k$ and $f_{km}$,
which are respectively the so-called data patterns for the signal
state and the probe states \cite{data patterns 2010,data patterns
2013,spiral,oxford,paderborn}. The goal of the data-pattern
reconstruction scheme is to find the \emph{estimated}
coefficients $c_m$ that give the closest match between the signal
patterns \{$F_k$\} and the probe patterns \{$f_{km}$\}. In all
previous studies on the data-pattern scheme, this matching
(or fitting) was done with least-squares inversion by
minimising the squared distance
\begin{equation}
\label{distance1}
D=\sum\limits_{k=1}^M\left(F_k-\sum_{m=1}^M c_m f_{km}\right)^2
\end{equation}
for the estimated coefficients $c_m$. We need a positive
semidefinite state estimator $\rho_\text{est}\ge 0$ that is of
unit trace, where the latter linear constraint can be
straightforwardly imposed with a parametrization involving $M-1$
independent coefficients, such that $c_M = 1 - \sum_{m=1}^{M-1}
c_m$. Different versions of least-squares inversion that
account for the positivity constraint and other general
linear constraints were used for data-pattern matching
\cite{data patterns 2010,data patterns 2013}). Efficient inference
procedure that directly incorporates the positivity constraint
in the search algorithm that improves the data pattern fit
by accumulating more data patterns from a pre-chosen set was
also established in Ref.~\cite{spiral}.

With the basic mathematical formalism now in place, one can
understand why it is generally unnecessary to accumulate a huge
dataset from many measurement settings to reconstruct unknown
signals. For instance, in the idealized hypothetical situation
where statistical fluctuation is absent, if the signal state is
any one of the probe states, say $\rho=\rho_{m=m_0}$,  then one
measurement setting ($\Pi_1$) is enough to determine the state
since one of the probe patterns $f_{1m_0}$ would precisely match
the signal data pattern $F_1$. All other probe patterns give
nonzero differences with $F_1$. This simple intuition suggests
that in a realistic scenario where statistical fluctuation is
present, if the unknown signal state is a linear combination of a
few of the probe states, there should exist a numerical method to
systematically search for a small set of optimized measurement
settings that would typically be much less than the total number
of probe states considered. This significant reduction in the total number $K$ of
optimal measurement settings required would also greatly enhance the numerical
efficiencies in reconstructing the state. In the
discussions to come, for a fixed set of probe states, we shall
derive a recursive iterative algorithm that decisively selects
optimal measurement settings in a sequential manner using Bayesian
statistical reasonings for finite data.

To choose optimal measurements sequentially, we shall
introduce a numerical procedure that invokes an iterative
Bayesian-update routine on the probability distribution for
the state estimator that is a function of the coefficients
$c_m$. The observer would start with a large bank of data
patterns that are acquired through different measurement settings
prior to the signal reconstruction. Since all information about
the measurement set-up is encoded in the data patterns, the
frequencies of the measurement outcomes, along with any
existing systematic errors, for the unknown signal state are
also encoded in its data patterns $F_k$. As this iterative
algorithm looks for optimal solutions by inferring from data
patterns, all systematic errors are automatically accounted
for.

Let us discuss the essential elements for the iterative
algorithm. After collecting data for $k$ measurement settings, the
corresponding \emph{posterior probability distribution}
$w\left(\vec{c}\, |\vec{F}^{(k)}\right)$ describes the
distribution of the column of coefficients $\vec{c}=(c_1 \ldots
c_{M-1})^\textsc{T}$, conditioned on the signal data patterns
$\vec{F}^{(k)}=(F_1 \ldots F_{k})^\textsc{T}$ obtained with $k$
different measurements. Performing an additional measurement of
a different setting enlarges the set of data patterns for the
signal. Consequently, the updated posterior distribution is
\begin{widetext}
\begin{eqnarray}
\begin{aligned}
\label{eqn:posterior_distr}
w\left(\vec{c}\, \Big|\vec{F}^{(k+1)}\right) =
\frac{w\left(\vec{c}\, \Big|\vec{F}^{(k)}\right)[P_{k+1}(\vec{c})]^{NF_{k+1}}\left[1-P_{k+1}(\vec{c}) \right]^{N(1-F_{k+1})}}
{\int d\vec{c}\, w\left(\vec{c}\, \Big|\vec{F}^{(k)}\right)[P_{k+1}(\vec{c})]^{NF_{k+1}} \left[1-P_{k+1}(\vec{c}) \right]^{N(1-F_{k+1})}},
\end{aligned}
\end{eqnarray}
\end{widetext}
where the estimated probability for the $k$th measurement setting reads
\begin{equation}P_{{k}}(\vec{c})\approx \sum_{m=1}^{M-1} c_m f_{{k}m}+\left(1-\sum_{m=1}^{M-1} c_m\right)f_{kM},
\label{faked probability}
\end{equation}
the integration measure is given by
\[d\vec{c}=\prod\limits_{m=1}^{M-1}dc_m,
\]
and $N$ is the number of signal-state copies used for each
measurement setting. The integration in the denominator of
Eq.~(\ref{eqn:posterior_distr}) is carried out over the
region $\Omega_c$ of coefficients $c_m$ where the estimator for $\rho$ in (\ref{eqn:density_matrix}) is positive
semidefinite.

\section{Quantum complexities in Bayesian reasoning}
\label{practicalities}

Generally, performing Bayesian updates for an unknown state
characterized by a large number of state parameters is
computationally demanding, since it involves calculating
operator integrals over the region of admissible parameter
values. (In our context, the denominator in
Eq.~(\ref{eqn:posterior_distr}).) This region is defined by
the positivity constraint of the state estimator,
$\rho_\text{est}\ge 0$, with the boundary representing
rank-deficient states. To estimate the integral in
Eq.~(\ref{eqn:posterior_distr}), a number of statistical
Monte-Carlo methods were suggested and implemented (see, for
example, \cite{robin2010} and the relevant references therein).
Incidently, the Bayesian procedure for choosing the measurements depending on the collected data was proposed in
Ref.~\cite{husar}. This procedure also involves a statistical
sampling method (sequential importance sampling \cite{douche})
that still does not effectively sample the state space according
to the required posterior distribution. Only recently, a more
efficient and direct Monte-Carlo approach that invokes Hamiltonian
statistical methods to sample the state space according to any
given arbitrary posterior distribution was introduced in
Ref.~\cite{berge}. Despite this breakthrough, while this approach
is efficient for single- and two-qubit cases, Hamiltonian
Monte-Carlo sampling for Hilbert spaces of larger dimensions
remains to be a relatively formidable task that requires more
computational resources and a cleverer operator-space
parametrization. For practical purposes, we shall henceforth
adopt a much simpler and far more computationally efficient
approximation that is similar to Kalman filtering (see
\cite{kalman} and the relevant references therein). As it will
be seen below, it follows naturally from the specifications of
the data-pattern reconstruction scheme.

First, let us outline some reasonable and useful approximations
that are in accordance with the specifications of the
data-pattern scheme. For a finite number of state copies
$N$, the probe patterns $f_{km}$ are random variables. We
shall assume that $N$ is sufficiently large, so that the
maximum of the (Gaussian-approximated) posterior distribution $w\left(\vec{c},|\vec{F}^{(k)}\right)$ corresponds to a positive semidefinite $\rho$. This assumption is justified when the
signal state is inside the state space (full rank) --- the real
situation in practice --- and its consideration in practical
implementations of the data-pattern reconstruction scheme 
on quantum states of the electromagnetic field can be
appreciated in \cite{oxford,paderborn}. Measuring large samples
of quantum systems in these experimental situations is typically
not a very resource-intensive task.

Next, we proceed, in the spirit of Kalman filtering, to
approximate the posterior distribution
$w\left(\vec{c},|\vec{F}^{(k)}\right)$ with a Gaussian
distribution of a data-dependent mean and variance (see
Ref{s}.~\cite{kalman} and \cite{robin2010}). Thus, assuming the
uninformative uniform prior distribution in the
$\{c_m\}$-space, we arrive at the following simplified form of
Eq.~(\ref{eqn:posterior_distr}):
\begin{widetext}
\begin{eqnarray}
\label{eqn:diag:posterior_distr}
w\left(\vec{c}\, \Big|\vec{F}^{(k+1)}\right)  {\propto}\,w\left(\vec{c}\, \Big|\vec{F}^{(k)}\right)
\exp\left( - \sum_{m,n=1}^{M-1} \frac{[f_{k+1,m}-f_{k+1,M}][f_{k+1,n}-f_{k+1,M})]}{2 \sigma_{k+1}^2}c_m c_n\right)\\
\nonumber
 \times \exp\left(\sum_{m=1}^{M-1} \frac{ [\mu_{k+1}-f_{k+1,M}][f_{k+1,m}-f_{k+1,M}] }{\sigma_{k+1}^2} c_m\right),
\end{eqnarray}
\end{widetext}
where
\begin{eqnarray}
\label{eqn:diag:sigma_mu} \sigma_{k+1}^2 = \frac{(NF_{k+1})
(N(1-F_{k+1})+1)}{(N+2)^2 (N+3)}, \\
\nonumber
\mu_{k+1} =
\frac{NF_{k+1}+1}{N+2}.
\end{eqnarray}
Here, Eq.~(\ref{faked probability}) is used for evaluating $P_{k+1}(\vec{c})$.

Notice that one can cope with a posterior distribution
extending outside the state space by using the procedure
outlined in Ref.~\cite{kalman}. The procedure involves
calculating the mean and variance for only the truncated
posterior distribution of Eq.~(\ref{eqn:diag:posterior_distr})
that lies in the state space, and redefine the posterior
distribution in the next step of the iteration to be the
Gaussian distribution having these distribution parameters,
so that no more than a very small portion of the updated
posterior distribution extends outside the physically allowed
space, or
\begin{equation}
\label{rematch}
\int_{\substack{\text{outside the}\\ \text{state space}}} d\vec{c}\,w\left(\vec{c}\,\Big|\vec{F}^{(k)}\right)\leq \epsilon
\end{equation}
for some $0<\epsilon\ll 1$. More details are given in the Appendix.

\section{Choosing optimal measurement settings}
\label{choice}

To optimally select the appropriate subset of measurement
settings out of all the choices that were used to accumulate the
bank of data patterns prior for signal reconstruction, a recursive
scheme to choose the next measurement setting according to the
previously used settings and a suitable criterion for terminating
this scheme are in order.

To set the stage for implementing the algorithm, we first specify
that in this article, we shall take the mean squared error
\begin{equation}
\Delta_\rho^2=\sum^{M-1}_{m=1}\left<\left(c_m-c^{(\text{true})}_m+\{\text{noise}\}\right)^2\right>_w
\label{eq:mean_sq_err}
\end{equation}
to be the measure of the (averaged) reconstruction accuracy of the
state estimator $\rho_\text{est}=\sum_mc_m\rho_m$ relative to the
true signal state $\rho$, where $\langle c_m \rangle_w = \int
d\vec{c} \; w\left(\vec{c}\, \Big|\vec{F}^{(k)}\right) c_m$ is the
mean with respect to the posterior distribution of the $c_m$s and
\{noise\} refers to an unbiased random perturbation on the signal
state $\rho$. Expanding the right-hand side gives
\begin{equation}
\Delta_\rho^2=\mathrm{Bias}[\{c_m\}]+\mathrm{Var}[\{c_m\}]+\sigma^2_\mathrm{noise}\,,
\end{equation}
where the magnitude of the bias term
\begin{equation}
\mathrm{Bias}[\{c_m\}]=\sum^{M-1}_{m=1}\left(\langle c_m \rangle_w-c^{(\text{true})}_m\right)^2
\end{equation}
depends on the choice of reconstruction scheme for translating
measurement data to a quantum state and the magnitude of
experimental systematic errors. The size of
$\sigma^2_\mathrm{noise}$ depends on the random environmental
noise. Since we shall only consider, in the hypothetical absence
of the positivity constraint and systematic errors, estimators
$\rho_\text{est}$s that are generated from the data in an unbiased
way $\left(\langle c_m \rangle_w=c^{(\text{true})}_m\right)$, the
bias term becomes insignificant in the situation of large $N$,
negligible systematic errors and realistic sources described by
full-rank signal states, albeit very close to the boundary for
practical quantum protocols. The noise variance also vanishes when
the source is relatively well stabilized. As a result, only the
variance term
\begin{equation}
\mathrm{Var}[\{c_m\}]=\sum^{M-1}_{m=1}\left<(c_m-\left<c_m\right>_w)^2\right>_w
\end{equation}
is relevant in determining the estimator's accuracy. One can thus
appreciate that when changes in the variance for each
subsequent iterative step become smaller than the variance
itself, then carrying out these subsequent steps by incorporating
additional measurement settings into the reconstruction will not
improve the reconstruction accuracy.

In view of the above reasoning, the natural figure of merit that
reflects our knowledge about the investigated unknown signal
state during the $k$th step of the iterative procedure, as
a function of $ w\left(\vec{c}\, \Big|\vec{F}^{(k)}\right)$,
would be the coefficient variance:
\begin{align}
\label{eqn:variance} \operatorname{Var}(\vec{c}|w,k) &\equiv{\mathrm{Var}[\{c_m\}]}\nonumber\\
&=\sum_{m=1}^{M-1} \left\langle \left(c_m - \langle c_m \rangle_w
\right)^2 \right\rangle_w\\
\nonumber &=\int d\vec{c} \; w\left(\vec{c}\, \Big|\vec{F}^{(k)}\right)
\sum_{m=1}^{M-1}(c_m-\langle c_m \rangle_w)^2.
\end{align}

The estimated information gained by incorporating the $(k+1)$th
measurement is therefore manifested as a decrease in the
coefficient variance for this measurement. The predicted
average variance after the ($k+1$)th measurement is then
given by the average over all possible data of a given $N$:
\begin{eqnarray}
\label{eqn:predicted_variance} \langle
\operatorname{Var}(\vec{c}|\Pi_{k+1})\rangle = \sum_{n=0}^N
p\left\{w;n,k\right\}\times\;\\
\nonumber
 \operatorname{Var}(\vec{c}|w,k+1;
F_{k+1} = n / N),
\end{eqnarray}
where $p\left\{w;n{,k}\right\}$ is the probability of obtaining
$n$ successful outcomes of $\Pi_{k+1}$;
$\operatorname{Var}(\vec{c}|w,k+1; F_{k+1} = n / N)$ is calculated
from Eq.~(\ref{eqn:variance}) for $\vec F^{(k+1)}  = (F_1, \ldots,
F_k, F_{k+1} = n / N)$ (thus, all possible measured frequencies
$F_{k+1} = 0, 1/N,\ldots, 1$ are included in average variance
estimation). This probability is estimated based on the previous
posterior distribution $w\left(\vec{c}\,
\Big|\vec{F}^{(k)}\right)$:
\begin{align}
\label{eqn:expected_probability}
&p\left\{w;n{,k}\right\}\\
\nonumber
=&\int d\vec{c} \;\binom{N}{n} [P_{k+1}(\vec{c})]^{n}
\left[1-P_{k+1}(\vec{c})\right]^{N-n} w\left(\vec{c}\,
\Big|\vec{F}^{(k)}\right).
\end{align}

So, according to our prescription, for the ($k+1$)th step of the
iterative update procedure, the observer should choose the measurement setting that
minimizes the predicted average variance, that is one should look
for
\begin{equation}
\Delta_{k+1}=\min_{\substack{\text{all available}\\ \text{settings}}} \Big\{\langle \operatorname{Var}(\vec{c}|\Pi_{k+1})\rangle\Big\},
\label{minimal variance}
\end{equation}
{where the minimum is carried out over} all {measurement settings
from the data-pattern bank}. {Hence, the criterion for terminating
the iteration is when} the {decrease in} the {predicted} variance
is much less than the variance {itself},
\begin{equation}
|\Delta_{k+1}-\operatorname{Var}(\vec{c}|w,k)|\ll \operatorname{Var}(\vec{c}|w,k),
\label{stopping rule}
\end{equation}
where $\Delta_{k+1}$ is given by Eq.~(\ref{minimal variance}).

One can surmise that for some
signal states (for example, the probe states themselves) only a few
iterative steps (measurements on the signal) would be sufficient to
obtain a good fit of the posterior distribution. In the preceding section, we shall demonstrate
that this is indeed the case for various kinds of signals.

We close this section by reiterating that it is possible to
technically cope with the numerical iteration even when the tails
of the posterior distribution typically extends to regions outside
the state space. One can improve on the final estimation of
the posterior distribution by truncating the part outside the
state space, followed by a Gaussian refitting. If the change in
the variance becomes negligible, then approximating the
correct posterior distribution with a function that extends
outside the state space will not lead to any appreciable
difference as far as signal reconstruction is concerned.

\section{Discussions \& examples}
\label{examples}

In this section, we illustrate the mechanisms and results of the
full recursive data-pattern tomography scheme. To this end, we
shall specify the set of probe states and measurements for this
scheme.

\begin{figure}[t]
\centering
\includegraphics[width=1\columnwidth]{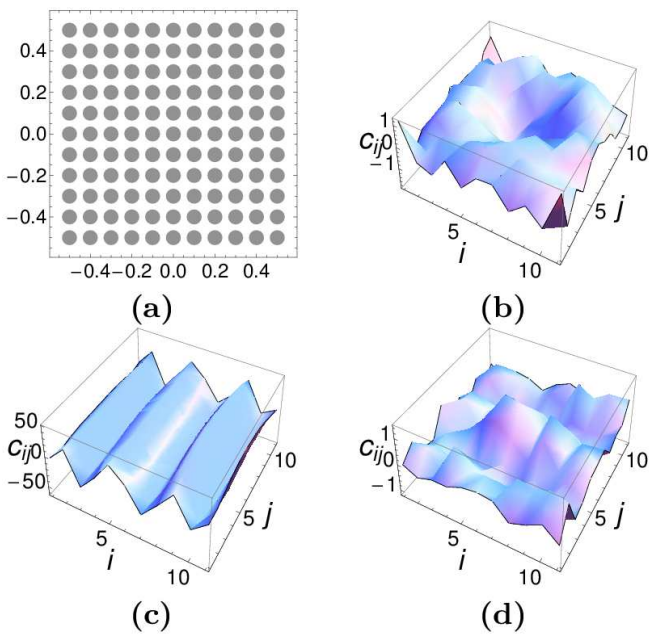}
\caption{(Color online) (a) The lattice of probe coherent states
in phase-space representation, indicated by an array of gray
filled circles. The abscissa represents the real part of the
complex amplitude that defines the phase space, whereas the
ordinate represents the imaginary part. (b) Coefficients $c_m$
for the single-photon signal. (c) Coefficients $c_m$ for the
signal coherent state with an amplitude $\alpha=1$ which
magnitude is larger than the lattice width. (d) Coefficients
$c_m$ for the signal even coherent state,
$|\psi\rangle\propto|\alpha=0.5\rangle+|\alpha=-0.5\rangle$.
Indices $i,j$ label the points on the lattice in panel (a);
$i$ labels the abscissa starting from the lower-left corner,
$j$ labels the ordinate starting from the lower-left corner.}
\label{fig1}
\end{figure}

\begin{figure}[t]
\centering
\includegraphics[width=0.9\columnwidth]{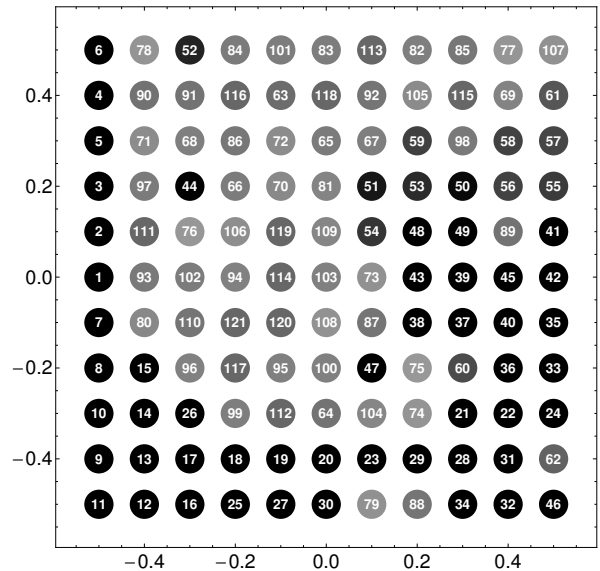}
\caption{Simulation of the reconstruction process for the coherent
state of amplitude $\alpha=0.5$, which is one of the probe states
employed. Phase-space representation of all available measurement
settings, where the lattice circles in Fig.~\ref{fig1}(a) are
replaced by filled numbered circles to display the sequence of
optimal measurement settings as chosen by the recursive Bayesian
algorithm. Black circles represent measurement-optimization stages
before the stopping criterion is satisfied, and circles of shades
of gray represent stages after the criterion is satisfied, with
the lightest shade indicating the smallest change in
reconstruction accuracy relative to the preceding step.} \label{fig20}
\end{figure}

\begin{figure}[t]
\centering
\includegraphics[width=\columnwidth]{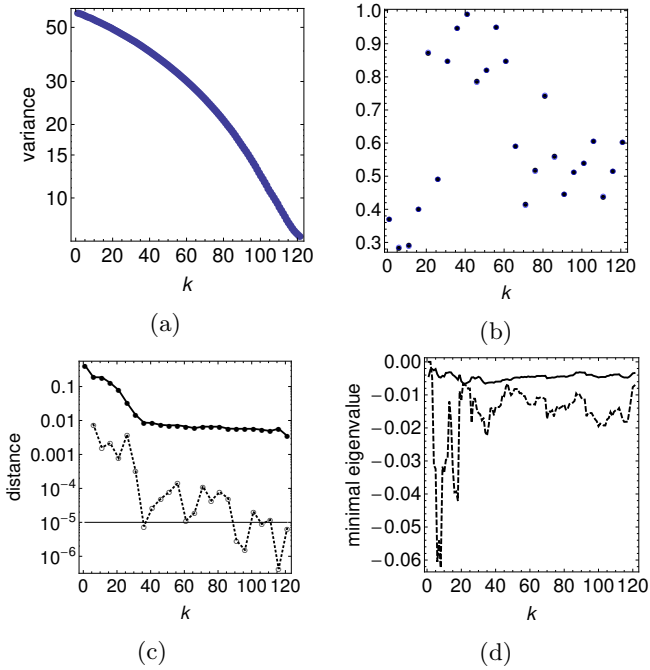}
\caption{(Color online) Simulation of the reconstruction process
for the coherent state of amplitude $\alpha=0.5$, which is one
of the probe states employed.  (a)~Plot of the variance
$\langle\operatorname{Var}(\vec{c}|\Pi_{k})\rangle$ against the
iterative step number or the optimal setting label.
(b)~Reconstructed probabilities of measurement settings for
the state estimator (filled circle) in the last step of the
Bayesian iteration (blue empty circles) and the corresponding
measured frequencies (black filled circles that are connected
to the blue ones, both types of which in this case are close to each other). (c)~Distance between the true state and
its estimator averaged over the posterior distribution (solid
lines and filled circles); distance between estimated signal
states (averaged over the posterior distribution) for the $k$th
and $(k+1)$th iterative steps (dotted line and empty circles).
The number of copies per measurement setting is $N=1000$ for all
probe and signal states. (d)~The minimal eigenvalues for the
estimators before (dotted line)
and after shearing-refitting (solid line) for the $k$th iterative step.} \label{fig2}
\end{figure}

\subsection{{Probe states}}

For practical purposes, it is advisable to use probe
states that can be easily generated in the laboratory, and
provide an accurate and computationally manageable representation
of the unknown signal state. In previous works on the
data-pattern scheme, it was shown that the usual coherent
states satisfy these practical requirements rather well. One can
routinely generate a large set of coherent states that can
potentially provide accurate representations for a wide class of
signal states \cite{data patterns 2010,data patterns 2013}. Apart
from coherent states, one can also take a general set of Gaussian
states as suitable candidates for the probe states. An example
would be a set of thermal mixtures of coherent states.
Interestingly, it has been shown that the inclusion of just one
such thermal state in the set of probe states involving other
Gaussian states can lead to a significant reduction in the
number of probe states required to accurately represent some
signal states \cite{data patterns 2013}.

To proceed we consider a set of probe coherent states that
defines a square lattice that is centered at the origin of
the two-dimensional phase space (see Fig.~\ref{fig1}(a)).
Without loss of generality, we limit ourselves to signal
states that are characterized by small amplitudes in phase
space. For the subsequent simulation examples, we consider three
different signal states: the coherent state with amplitude
$\alpha=1$, the single-photon Fock state, and the even coherent
state of amplitude $\alpha=0.5$, that is the superposition
$|\psi\rangle\propto|\alpha=0.5\rangle+|\alpha=-0.5\rangle$.
Figures~\ref{fig1}(b) through \ref{fig1}(d) illustrate these
three signal states in phase space. With this set of probe
states, the optimal fitting for all three signal states
respectively give the operators $\rho$ that have essentially 100\%
fidelity with the signal states, which are exemplifying
indications that our chosen probe-coherent-state basis can
reliably represent a wide class of signal states.

\begin{figure}[t]
\centering
\includegraphics[width=0.9\columnwidth]{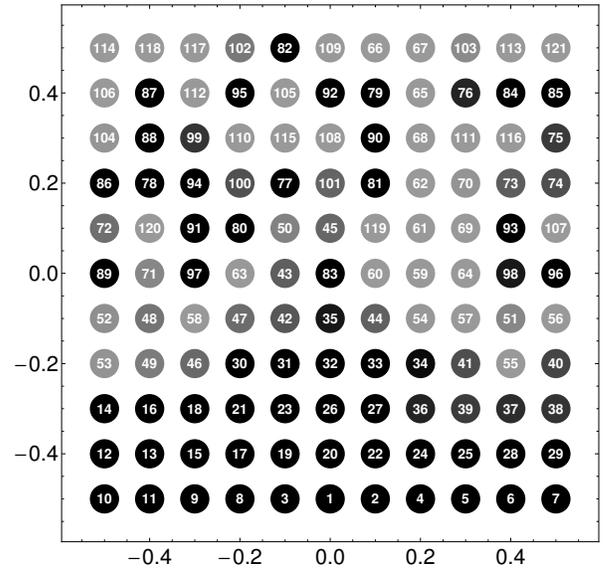}
\caption{Simulation of the reconstruction process for the
single-photon state. See Fig.~\ref{fig20} for the description of
the figure.} \label{fig30}
\end{figure}

\begin{figure}[t]
\centering
\includegraphics[width=\columnwidth]{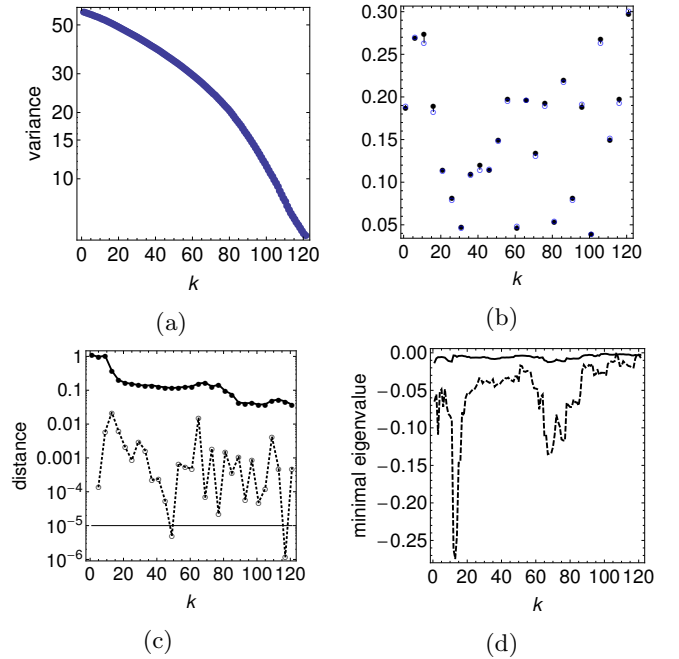}
\caption{(Color online) Simulation of the reconstruction process
for the single-photon state. See Fig.~\ref{fig2} for the
descriptions of all figure panels.} \label{fig3}
\end{figure}

\subsection{Phase-Space Sampling}
\label{subsec:PSsampling}

As a means of showcasing the mechanism of our recursive
Bayesian method in the simplest and most straightforward
way, let us take the intended measurements to be
projections onto coherent states, so that the $k$th
measurement is described by observing a coherent-state outcome
$\Pi_k=|\beta_k\rangle\langle\beta_k|$ of some chosen complex
amplitude $\beta_k$. A collection of these measurement projections
define the scope of the two-dimensional phase space that is
sampled by these measurements. For the assumed ideal lossless
detection, the probability of observing the $k$th coherent-state
outcome given a probe state $\rho_m$ is given by
\begin{equation}
p_{km}=\langle\beta_k|\rho_m|\beta_k\rangle=\exp\!\left(-|\alpha_m-\beta_k|^2\right),
\label{probability probes}
\end{equation}
where $\alpha_m$ denotes the complex amplitude of the probe
coherent state. Deterministic sampling of the phase space is achievable experimentally using the unbalanced homodyne technique \cite{unb_homo1,unb_homo2} and has, for instance, been applied to wave-function reconstruction \emph{via} phase retrieval \cite{unb_homo3} and Bell-inequality testing \cite{unb_homo4}. For non-ideal detection, this method of
systematically sampling the phase space has been demonstrated
experimentally \cite{paris}.

For numerical experiments simulated with Monte~Carlo techniques,
the probabilities of respectively observing every coherent-state
setting $k$ for the $m$th probe state are measured with a
finite number of state copies $N_{\text{p}}$. The frequencies
are taken as the probe patterns. Amplitudes of the probe
states are selected as equidistant phase-space points that form
a square lattice (see Fig.~\ref{fig1}(a), for instance). Note
that the matrix of the probabilities $p_{km}$ is generally not
square. Typically, the number of employed measurement settings $K$ is
not equal to the number of probe states $M$, and the sets of
amplitudes \{$\alpha_m$\} and \{$\beta_k$\} may not
even be overlapping in the sense that $\alpha_m\neq\beta_k$
for all $k$ and $m$. However, for the sake of simplicity we
let $M=K$ and $\alpha_m =\beta_m$. We also assume that the
signal is measured for a finite number of state copies
$N_{\text{s}}$ per measurement setting such that
$N_{\text{p}}=N_{\text{s}}=N$. Prior to the reconstruction,
the probe-pattern bank was obtained by simulating all the
projections on the phase-space lattice for every probe state with
$N=1000$ copies.

\subsection{Results}
\label{subsec:results}
We are now ready to demonstrate that the recursive Bayesian
iterative algorithm defined in Sec.~\ref{choice} can significantly
reduce the number of optimal measurements needed for a faithful
reconstruction of the signal state. For this purpose, we take the
signal state to be one of the probe coherent state with
amplitude 0.5. Figures~\ref{fig20} and \ref{fig2} illustrate aspects
of a simulation conducted with 121 probe states residing in the
$11\times11$ phase-space square lattice (see Fig.~\ref{fig20}), which is also the same lattice illustrating
the measurement settings. Figure~\ref{fig20} highlights the
sequence of optimal measurement settings dictated by the recursive
Bayesian algorithm as phase-space trajectories (indicated by
integers in filled circles). The evolution phase-space trajectory
is self-explanatorily indicated by the numbered circles.
Figure~\ref{fig2}(a) gives the plot of total variance of
the posterior distribution,
$\langle\operatorname{Var}(\vec{c}|\Pi_{k})\rangle$, against the
iterative step number (with 121 being the largest) or the optimal
setting label. From Fig.~\ref{fig2}(c), one can see that
after measuring around 58 optimal measurement settings, the
changes in the squared distances, defined by
Eqs.~\eqref{eqn:variance} and \eqref{eqn:predicted_variance},
become much smaller than the distances themselves, implying
that no further improvements can be exploited. One can also look
at the distance between estimated signal states (averaged
over the posterior distribution) for the $k$th and
$(k+1)$th optimal setting (dotted line in
Fig.~\ref{fig2}(c)). Figure~\ref{fig2}(d) illustrates how the shearing-refitting procedure affects the minimal eigenvalue of the
estimated signal state.

\begin{figure}[t]
\centering
\includegraphics[width=0.9\columnwidth]{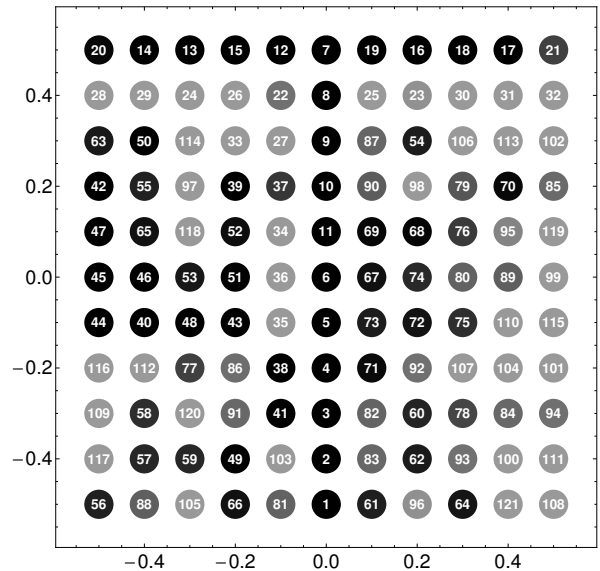}
\caption{(Color online) Simulation of the reconstruction process
for the even coherent state of amplitude $\alpha=0.5$. See
Fig.~\ref{fig20} for the description of the figure.}
\label{fig40}
\end{figure}

\begin{figure}[t]
\centering
\includegraphics[width=\columnwidth]{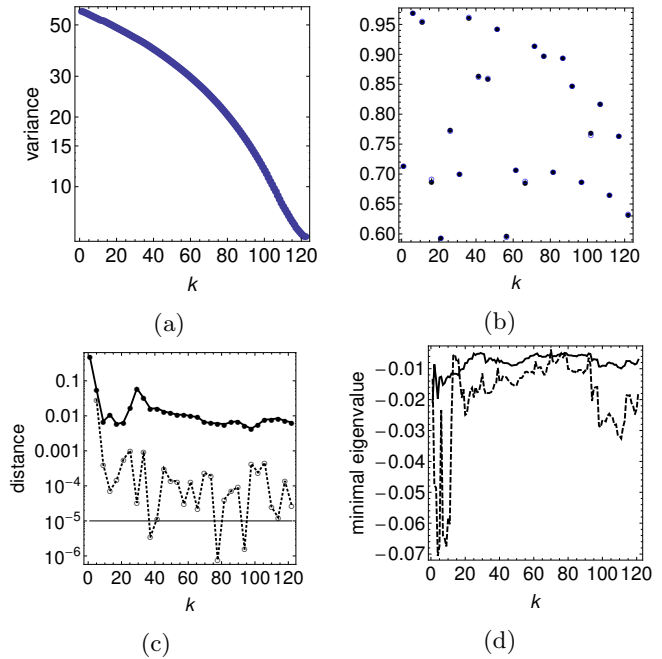}
\caption{(Color online) Simulation of the reconstruction process
for the single-photon state. See Fig.~\ref{fig2} for the
descriptions of all figure panels.} \label{fig4}
\end{figure}

Figures~\ref{fig30} and \ref{fig3} show the simulated
reconstruction process for a single-photon state, which is a
nonclassical state having a highly singular Glauber--Sudarshan P~function.
All the typical features of the iterative procedure for the
coherent-state signal are also present here. Again, the
total number of optimal measurement settings employed is less than the number of parameters
$M$ (Reading off from the panels (b) and (c), about $K=100$
measurement settings is sufficient for an accurate signal
reconstruction). For the same $N$ value and stopping criterion,
the reconstruction accuracy of the estimator for the coherent
state is higher than that of the estimator for the single-photon
state.

Figures~\ref{fig40} and \ref{fig4} pertain to the reconstruction
of another nonclassical state: the even coherent state
$|\psi\rangle\propto|\alpha=0.5\rangle+|\alpha=-0.5\rangle$.
One can see that, again, all the prominent features of the
reconstruction procedure are the same as in the previous two
examples. The number of optimal measurement settings needed (about $K=90$) is also
less than the total number of parameters.

The reconstruction accuracy of the estimator for the even
coherent state is higher than the single-photon state, as can be
clearly seen by comparing the (b) panels in Figs.~\ref{fig3}
and \ref{fig4}, where the estimated probabilities and
measured frequencies are shown. On hindsight, it is to be
expected that the single-photon state results in a less faithful
representation in the coherent-state basis relative to the
even (or odd) coherent state. The reason is that this state has a rather large classical component. An equal statistical mixture of just two coherent projectors of respectively $\ket{\alpha=0.5}$ and $\ket{\alpha=-0.5}$ would already give a rather high fidelity (0.8894).

We end this discussion by briefly summarizing the numerical
technique that enforces the localization of the posterior
distribution inside the admissible state-space. The update
formulas (\ref{faked probability}) and
(\ref{eqn:diag:posterior_distr}) essentially specify a Gaussian
distribution for the coefficient vector $\vec c$ without imposing
the positivity restriction. In each iterative step of the
reconstruction procedure, just before the estimation of the
average variance, a corrected Gaussian distribution that localizes
inside the state space is computed in the manner described in
Ref.~\cite{kalman}.

To incorporate the positivity constraint for quantum states,
we approximate the state space by a set of linear inequality
constraints on the coefficient vector $\vec{c}=(c_1\,\,c_2\,\,\ldots\,\,c_{M-1})^\textsc{T}$. According to
the definition of a quantum state, a density operator $\rho$
must satisfy the inequality $ \left\langle{\Psi} \left| {\rho}
\vphantom{\Psi}\right| \kern-\nulldelimiterspace
{\Psi}\right\rangle \ge 0$ for any ket
$\left|{\Psi}\right\rangle$. In practice, we can only account
for a finite set of inequality constraints $
\left\langle{\Psi_i} \left| {\rho} \vphantom{\Psi_i}\right|
\kern-\nulldelimiterspace {\Psi_i}\right\rangle \ge 0$ defined by
a finite set of test kets $\{|\Psi_i \rangle\}$. If this set of
test kets is large enough, the finite set of inequalities
provides a good approximation for the positivity constraint for
operator $\rho$. It is worth noting, that if $\rho$ has a
finite number of non-zero eigenvalues and the set of test kets
includes all the eigenkets of $\rho$ corresponding to non-zero
eigenvalues, the set of inequalities provide an exact
description of the state space. In the examples above,
the test-ket set includes the Fock states $|n\rangle$, with
$n=0,\ldots,40$, and the probe coherent states. Equation~\eqref{eqn:density_matrix} implies that the coefficients $c_m$
must satisfy the following inequalities
\begin{equation}
\label{eqn:linear_constraints}
\sum_{m=1}^{M-1} c_m v_{m,i} \ge u_i,
\end{equation}
where $ v_{m,i} = \left\langle{\Psi_i} \left| {\rho_m}
\vphantom{\Psi_i}\right| \kern-\nulldelimiterspace
{\Psi_i}\right\rangle - \left\langle{\Psi_i} \left| {\rho_M}
\vphantom{\Psi_i}\right| \kern-\nulldelimiterspace
{\Psi_i}\right\rangle $ and $u_i = - \left\langle{\Psi_i} \left|
{\rho_M} \vphantom{\Psi_i}\right| \kern-\nulldelimiterspace
{\Psi_i}\right\rangle $, in order to satisfy the positivity
constraint.

Appendix \ref{appendix} describes the iterative procedure of
distribution shearing of the initial posterior distribution with
a significant portion extending outside the state space for a
finite set of linear constraints, so that the final
(Gaussian) posterior distribution extends outside the state
space only minimally.

\section{Conclusions}
We have developed a simple and practical Bayesian
procedure for enhancing the data-pattern reconstruction of signal
states. The essence of the procedure is to sequentially choose the
minimal set of optimal measurements needed to faithfully reconstruct the signal
with a pre-chosen basis of probe states. In each step of the procedure, the
measurement setting that minimizes the estimation error specified by the posterior
distribution is incorporated into the previously measured settings. Upon consideration of specific features of the
data-pattern scheme, we establish this recursive Bayesian procedure in
the spirit of Kalman filtering. Such an approach allows
us to handle large probe-state bases.
We have demonstrated the mechanisms of our proposed Bayesian recursive data-pattern procedure with a set of more
than one hundred probe coherent states to obtain accurate and efficient reconstruction
of the coherent-state, single-photon-state and even-coherent-state signals.
For some signal states, our recursive procedure even allows for accurate reconstruction using considerably
less measurement settings than the elements of the probe-state basis used
to represent the state. We expect that our scheme will
be of much use in practical applications of data-pattern
tomography.

\section{Acknowledgments}
A.~M. and D.~M. acknowledge the support of the National Academy of Sciences
of Belarus through the program "Convergence", and FAPESP grant  2014/21188-0 (D.~M.). Y.~S.~T., J.~{\v R}., and Z.~H. acknowledge the support of the Grant Agency
of the Czech Republic (Grant No. 15-031945), the IGA Project of the Palacky University (Grant No. PRF
2015-002), and the European Union Seventh Framework Programme under Grant Agreement No. 308803
(Project BRISQ2).

\appendix
\section{Distribution shearing and Gaussian refitting}
\label{appendix}
The Gaussian posterior distribution can be parameterized as $w(\vec c) \propto \exp\left( - \vec c \cbdot \mat{A} {\cbdot}\vec c + {\vec b} \cbdot \vec c \right)$, where $\mat{A}$ is an $(M-1)\times(M-1)$ matrix and ${\vec b}$ is a column vector with $(M-1)$ elements. For convenience, we introduce the new vectorial variable
\begin{equation}
 \label{eqn:c'}
 \vec c' = \mat{A}^{1/2} {\cbdot} \vec c - \frac{1}{2} \mat{A}^{-1/2} {\cbdot\vec b},
\end{equation}
in which the expression for the posterior distribution simplifies immensely to $w(\vec c') \propto \exp\left(-\vec c' \cbdot \vec c' \right)$.
If the column $\vec c$ satisfies Eq.~\eqref{eqn:linear_constraints}, then the new column $\vec c'$ will satisfy a similar set of linear constraints $\vec c' \cbdot \vec v_i' \ge u_i'$, where
\begin{equation}
\label{eqn:vu}
\vec v_i' =  \mat{A}^{-1/2} {\cbdot} \vec v_i,\; u_i' = u_i - \frac{1}{2}\vec v_i \cbdot \mat{A}^{-1} {\cbdot} {\vec b}.
\end{equation}

If we had only one constraint, say $\vec c' \cbdot \vec v_1' \ge u_1'$, distribution shearing could be carried out in the following way. Owing to rotational symmetry of the Gaussian posterior distribution $w(\vec c')$, only variations along the direction of $\vec v_1' / \left| \vec v_1' \right|$ matters. Upon introducing another variable $x = \vec c' \cbdot \vec v_1' / \left| \vec v_1' \right|$, we consider the marginal distribution $w(x)\propto \exp(-x^2)$ with the constraint $x \ge x_0 = u_1' /  \left| \vec v_1' \right|$. Hence, distribution shearing and Gaussian refitting would yield a new marginal distribution $\tilde w(x)\propto \exp\left(-(1+a)x^2 + b x\right)$ of real coefficients $a$ and $b$ that gives the same mean value $\langle x \rangle$ for the physical portion of the initial distribution $w(x)$ inasmuch as
\begin{equation}
\label{eqn:constrained_mean}
\frac{\int_{x_0}^{\infty}{dx}\,xw(x)}{\int_{x_0}^{\infty}{dx}\, w(x)} = \frac{\int_{x_0}^{\infty}{dx}\, x \tilde w(x)}{\int_{x_0}^{\infty}{dx}\, \tilde w(x)},
\end{equation} but {with a} smaller probability
\begin{equation}
\label{eqn:probability_constrained}
p = \int_{-\infty}^{x_0}{dx}\, \tilde w(x)
\end{equation}
of the variable $x$ lying outside the physical region that is defined by its linear constraint. After some straightforward evaluations of the integrals in Eqs.~\eqref{eqn:constrained_mean} and \eqref{eqn:probability_constrained}, it can be shown that the coefficients $a$ and $b$ satisfy the following system of equations:
\begin{align}
\label{eqn:ab}
b &= 2 (1+a) x_0 - 2 \sqrt{1+a} \operatorname{erf}^{-1}(2p-1),\nonumber\\
\frac{\exp(-x_0^2)}{\sqrt{\pi} (1-\operatorname{erf}(x_0))} &= \frac{b}{2(1+a)} + \frac{ \exp\left[-\left(\operatorname{erf}^{-1} (2p-1) \right)^2\right] }{ 2 \sqrt{\pi} \sqrt{1+a} (1-p)}.
\end{align}
This linear system has a unique solution for ${a=}a(x_0{,p})$, ${b=}b(x_0{,p})$ that can be found easily using any sort of efficient numerical routines for solving equations. It then follows that the new Gaussian posterior distribution in $\vec c'$ takes the form
\begin{equation}
\label{eqn:constrained_distr'}
\begin{gathered}
\tilde w(\vec c') \propto\exp \Bigl[ -\vec c' \cbdot \vec c' - a(x_0{,p}) \left(\vec c' \cbdot \vec v_1' / \left|\vec v_1' \right|\right)^2 + \\ + b(x_0{,p}) \left(\vec c' \cbdot \vec v_1' / \left|\vec v_1' \right|\right) \Bigr].
\end{gathered}
\end{equation}
After reverting to the original column $\vec c$, we obtain
\begin{equation}
\label{eqn:constrained_distr}
\tilde w(\vec c) \propto\exp\left( - \vec c \cbdot (\mat{A} + \delta \mat{A}){\cbdot} \vec c + ({\vec b} + \delta {\vec b}) \cbdot \vec c \right),
\end{equation}
where
 \begin{equation}
 \label{eqn:newA}
 \delta \mat{A} = a(x_0{,p}) \vec v_1 {{\vec v_1}^\textsc{T}} / \left|\vec v_1' \right|^2
\end{equation}
and
\begin{equation}
\label{eqn:newB}
\delta {\vec b} = b(x_0{,p}) \vec v_1 / \left|\vec v_1' \right| + a(x_0{,p}) \left( \vec v \cbdot \mat{A}^{-1}{\cbdot} {\vec b} \right) \vec v / \left|\vec v_1' \right|^2.
\end{equation}

Our present situation, on the other hand, requires us to deal
with a set of more than one linear constraints. Naively, one
may attempt to carry out distribution shearing sequentially
for all the constraints. That is, one selects the first
constraint and calculates the quantities $\vec v_i'$ and $u_i'$
from Eq.~(\ref{eqn:vu}) and updates $\mat{A}$ and ${\vec b}$
according to Eqs.~(\ref{eqn:newA}) and (\ref{eqn:newB}), then
move on to the next constraint and repeat the previous procedures,
and so on. However, such a simplistic methodology usually
fails, since distribution shearing along one direction $\vec
v_i$ can make situation worse for several other directions $\vec
v_{j\neq i}$, thereby causing the procedure to diverge. An
important property of the elementary one-dimensional shearing
process, defined by Eqs.~(\ref{eqn:constrained_mean}) and
(\ref{eqn:probability_constrained}), is additivity: reducing
the violation probability from $p=p_0$ to $p=p_2$ is precisely
equivalent to first reducing it from $p=p_0$ to some intermediate
value $p=p_1$, followed by a second reduction from $p=p_1$ to
$p=p_2$. Therefore, the path that is taken to reduce the
violation probability $p$ is irrelevant. So, for a convergent
numerical method, in each iterative step of the shearing
process we choose the constraint with the strongest deviation
${\big|}x_0^{(i)}{\big|} = {\big|} u_i - \frac{1}{2}\vec v_i
\cbdot \mat{A}^{-1} {\cbdot\vec b}{\big|} / \left|
\mat{A}^{-1/2}{\cbdot} \vec v_i \right|$. The posterior
distribution $w(\vec c)$ is sheared and refit to $\tilde{w}(\vec
c)$ by considering this constraint, where $p$ for this
constraint is reduced by 0.25\% if it is greater than 1\%.
Otherwise, the shearing procedure terminates. For all the
examples in Sec.~\ref{subsec:PSsampling}, this method of
distribution shearing and Gaussian refitting converges rather
efficiently.

\end{document}